\documentclass[useAMS,usenatbib,usegraphicx]{mn2e}

\def\simlt{\lower.5ex\hbox{$\; \buildrel < \over \sim \;$}}
\def\simgt{\lower.5ex\hbox{$\; \buildrel > \over \sim \;$}}

\def\Lya{\mbox{Ly$\alpha$} }

\def\beq{\begin{equation}}
\def\eeq{\end{equation}}
\def\del{\delta}
\def\beqa{\begin{eqnarray}}
\def\eeqa{\end{eqnarray}}

\def\c2{c_{200}}

\def\r2{r_{200}}

\def\M2{M_{200}}

\def\zr{z_{\rm rei}}
\def\Vb{V_{\rm bub}}

\title{The Infancy of Cosmic Reionization} \author[Rennan
Barkana]{Rennan Barkana$^{1,2,3}$\thanks{E-mail:
barkana@wise.tau.ac.il}\\ $^{1}$ Institute for Cosmic Ray Research,
University of Tokyo, Kashiwa 277-8582, Japan \\ $^{2}$ Division of
Physics, Mathematics and Astronomy, California Institute of
Technology, Mail Code 130-33, Pasadena, CA 91125, USA \\ $^{3}$
Guggenheim Fellow; on sabbatical leave from the School of Physics and
Astronomy, Tel Aviv University, Israel}

\begin{document}

\pagerange{\pageref{firstpage}--\pageref{lastpage}} \pubyear{2008}
\maketitle
\label{firstpage}
\begin{abstract}
We consider the early stages of cosmic hydrogen or helium
reionization, when ionizing sources were still rare. We show that
Poisson fluctuations in the galaxy distribution substantially affected
the early bubble size distribution, although galaxy clustering was
also an essential factor even at the earliest times. We find that even
at high redshifts, a significant fraction of the ionized volume
resided in bubbles containing multiple sources, regardless of the
ionizing efficiency of sources or of the reionization redshift. In
particular, for helium reionization by quasars, one-source bubbles
last dominated (i.e., contained $90\%$ of the ionized volume) at some
redshift above $z=7.3$, and hydrogen reionization by stars achieved
this milestone at $z>23$. For the early generations of atomic-cooling
halos or molecular-hydrogen-cooling halos, one-source ionized regions
dominated the ionized volume only at $z>31$ and $z>48$,
respectively. To arrive at these results we develop a statistical
model for the effect of density correlations and discrete sources on
reionization and solve it with a Monte Carlo method.
\end{abstract}


\section{Introduction}\label{intro}

The earliest generations of stars are thought to have transformed the
universe from darkness to light and to have reionized and heated the
intergalactic medium. Knowing how the reionization process happened is
a primary goal of cosmologists, because this would tell us when the
early stars formed and in what kinds of galaxies. The strong
fluctuations in the number density of galaxies, driven by large-scale
density fluctuations in the dark matter, imply that the dense regions
reionize first, producing on large scales an inside-out reionization
topology \citep{BLflucts}. This basic picture has been studied and
confirmed with detailed analytical models \citep{fzh04},
semi-numerical methods \citep{mesinger}, and by a variety of large
numerical simulations \citep{mellema, zahn, cen} that solve gravity
plus radiative transfer. The distribution of neutral hydrogen during
reionization can in principle be measured from maps of 21-cm emission
by neutral hydrogen \citep{Madau}, although upcoming experiments such
as the Murchison Widefield Array
(MWA)\footnote{http://www.haystack.mit.edu/ast/arrays/mwa/} and the
Low Frequency Array (LOFAR)\footnote{http://www.lofar.org/} are
expected to be able to detect ionization fluctuations only
statistically \citep[for reviews see, e.g.,][]{fob06,bl07}.

The infancy of cosmic reionization, when only a small fraction of the
volume of the universe was ionized, is of interest for a number of
reasons. First, when ionizing sources were rare at early times, they
are expected to have formed separate H~II bubbles which if observed
can be used to study directly the properties of individual sources and
their surroundings \citep{cen2}, without the complications of later
times, when overlapping bubbles imply that galaxy clustering dominates
the ionization distribution and the 21-cm power spectrum.  Second,
when ionization fluctuations disappear over much of the universe, it
becomes possible to use the 21-cm technique for other applications
including those of fundamental cosmology, without the complications of
ionization fluctuations which are intrinsically non-linear (since the
ionization fraction varies from 0 to 1). Major such applications
include measurements of the density power spectrum
\citep{rees1,rees2}, of fluctuations in the \Lya radiation emitted by
the first galaxies \citep{BL05b,Jonathan06a,Shapiro}, and of
fluctuations in the rate of heating from early X-rays
\citep{Jonathan07}. If ionization fluctuations are negligible then the
angular anisotropy of the 21-cm power spectrum makes it possible to
measure separately various fluctuation sources, including in
particular the cosmologically-interesting baryonic density power
spectrum \citep{BL05a}. On small scales, the existence of H~II bubbles
(even when rare) affects the fluctuations in \Lya and X-ray radiation,
producing a small-scale cutoff in the 21-cm power spectrum that can be
used to detect and study the population of galaxies that formed just
200 million years after the Big Bang \citep{NB08}.

While analytical models and numerical simulations exist that can be
used to study the later epochs of reionization, the early times are
very difficult to investigate. Simulations, which in general must
overcome the huge disparity between the large characteristic scales of
galaxy clustering at high redshift and the small scales of individual
galaxies \citep{BLflucts}, are stretched even further at early times,
when ionizing sources become very rare and even larger cosmological
volumes are required in order to assemble a reasonable statistical
sample. As discussed in detail below, current analytical models based
on the model of \citet{fzh04} account for galaxy clustering but are
based on continuous variables and cannot account for the fact that
galaxies are discrete sources. This discreteness becomes a crucial
factor in the early stages of reionization, when the number of
ionizing sources per bubble is small. In this limit, Poisson
fluctuations also become substantial, weakening the correlation
between the galaxy distribution and the underlying large-scale density
fluctuations in the dark matter. Discreteness can also play a
significant role during the central stages of reionization,
particularly in the case of He reionization by quasars, which are rare
sources believed to form only in massive halos that correspond to
many-$\sigma$ density fluctuations at high redshift. These various
aspects of discrete sources are not accounted for in current
analytical models. \citet{FurOh} considered helium reionization and
showed that the continuous models break down when discreteness is
important. They suggested to instead use a pure stochastic Poisson
model, without halo correlations, when He is less than $\sim 50\%$
ionized globally.

In this paper we develop a model that accounts for discrete sources as
well as density correlations. We solve the model with a Monte Carlo
method and use it to show that galaxy correlations play a major role
even in the infancy of cosmic reionization. Isolated one-source
bubbles do dominate at sufficiently high redshifts, but the pure
stochastic Poisson model is essentially never a good description of
the bubble size distribution. In the next section we first review
previous models (section~\ref{s:prev}), then develop ours
(section~\ref{s:full}) and summarize all the various models whose
results we later compare (section~\ref{s:sum}). We illustrate our
results during the infancy of reionization (section~\ref{s:1perc}) and
then develop an approximate calculation that allows us to scan through
a wide parameter space of possible reionization scenarios
(section~\ref{s:aprx}). Finally, we illustrate our results during
later stages of reionization (section~\ref{s:late}) and summarize our
conclusions (section~\ref{s:conc}). We assume a standard $\Lambda$CDM
universe with cosmological parameters that match the five-year WMAP
data and other large scale structure observations \citep{wmap}, namely
$\Omega_m=0.28$ (dark matter plus baryons), $\Omega_\Lambda=0.72$
(cosmological constant), $\Omega_b=0.046$ (baryons), $h=0.7$ (Hubble
constant), $n=0.96$ (power spectrum index) and $\sigma_8=0.82$ (power
spectrum normalization).

\section{Model}

Analytical approaches to galaxy formation and reionization are based
on the mathematical problem of random walks with barriers. The
statistics of a random walk with a barrier can be used to calculate
various one-point distributions, including the distribution of ionized
bubble sizes during reionization \citep{fzh04}. This distribution
indicates how likely it is for each scale to determine whether a given
point is ionized. As such, it indicates the relative importance of
various scales in reionization, yielding important intuition about the
internal dynamics of reionization. If bubbles of a given radius $R$
are common, this produces a strong correlation in the neutral fraction
(and thus 21-cm emission) on a scale $\sim R$, since the ionization
states of two points separated by up to $R$ are then often
coupled. Calculations of the 21-cm correlation function using
two-point extensions of the model yield reasonable agreement with
numerical simulations \citep{fzh04,zahn,b07} and indicate that the
main feature of the power spectrum during reionization, i.e., enhanced
large-scale power, indeed appears on scales corresponding to the most
likely bubble sizes.

In this section we first review the basic setup of the random walk
problem in the context of reionization. We then show how the standard
approach can be generalized to solve for the bubble size distribution
including Poisson fluctuations.

\subsection{Reionization: basic setup}

\label{s:prev}

The basic approach for using random walks with barriers in cosmology
follows \citet{bc91}, who used it to rederive and extend the halo
formation model of \citet{ps74}. In this approach we work with the
linear overdensity field $\del({\bf x},z) \equiv \rho({\bf
x},z)/\bar\rho(z) - 1$, where ${\bf x}$ is a comoving position in
space, $z$ is the cosmological redshift and $\bar \rho$ is the mean
value of the mass density $\rho$. In the linear regime, the
overdensity grows in proportion to the linear growth factor $D(z)$
(defined relative to $z=0$), making it possible to extrapolate the
initial density field at high redshift to the present by
multiplication by the relative growth factor. Thus, in this paper the
density $\del$ and related quantities refer to their values
linearly-extrapolated to the present. In each application there is in
addition a barrier that signifies the critical value (as a function of
scale) which the linearly-extrapolated $\delta$ must reach in order to
achieve some physical milestone, which here corresponds to having a
sufficient number of galaxies within some region in order to fully
reionize it.

Considering an arbitrary point $A$ in space (at a given $z$), we
calculate as follows its probability of being inside H~II bubbles of
various sizes \citep{fzh04}. We consider the smoothed density around
this point, first averaging over a large scale or, equivalently,
including only small comoving wavenumbers $k$. We then average over
smaller scales (i.e., include larger $k$) until we find the largest
scale on which the averaged overdensity is higher than the barrier; in
the application to reionization, we then assume that the point $A$
belongs to an H~II bubble of this size. Mathematically, if the initial
density field is a Gaussian random field and the smoothing is done
using sharp $k$-space filters, then the value of the smoothed $\del$
undergoes a random walk as the cutoff value of $k$ is
increased. Instead of using $k$, we adopt the (linearly-extrapolated)
variance $S$ of density fluctuations as the independent
variable. While the solutions are derived in reference to sharp
$k$-space smoothing, we follow the traditional extended
Press-Schechter approach and substitute real-space quantities in the
final formulas. In particular, $S$ is calculated as the variance of
the mass $M$ enclosed in a spatial sphere of comoving radius $r$.

The appropriate barrier for reionization was derived by \citet{fzh04},
who noted that the ionized fraction in a region is given by $x^i =
\zeta F_{\rm coll}$, where $F_{\rm coll}$ is the collapse fraction
(i.e., the gas fraction in galactic halos) and $\zeta$ is the overall
efficiency factor, which is the number of ionizing photons that escape
from galactic halos per hydrogen atom (or ion) contained in these
halos, divided by the number of times each hydrogen atom in the
intergalactic medium must be reionized (where this number is assumed
to be spatially uniform). In the extended Press-Schechter model
\citep{bc91}, in a region containing a mass corresponding to variance
$S_R$, \beq F_{\rm coll} = {\rm erfc} \left(\frac{\del_c(z) - \del}
{\sqrt{2 (S_{\rm min} - S_R)}} \right) \ , \label{eq:exPS} \eeq where
$S_{\rm min}$ is the variance corresponding to the minimum mass
$M_{\rm min}$ of a halo that hosts a galaxy, $\del$ is the mean
density fluctuation in the given region, and $\del_c(z)$ is the
critical density for halo collapse at $z$. In reality, the cosmic mean
halo distribution in simulations is better described by the halo mass
function of \citet{shetht99} (with the updated parameters suggested by
\citet{st02}). However, an exact analytical generalization is not
known for the biased $F_{\rm coll}$ in regions of various sizes
(corresponding to $S_R$) and mean density fluctuations $\del$.

\citet{BLflucts} suggested a hybrid prescription that adjusts the
abundance in various regions based on the extended Press-Schechter
formula \citep{bc91}, and showed that it fits a broad range of
simulation results. In general, we denote by $f(\del_c(z),S)\, dS$ the
mass fraction contained at $z$ within halos with mass in the range
corresponding to variance $S$ to $S+d S$, where $\del_c(z)$ is the
critical density for halo collapse at $z$. Then the biased mass
function in a region of size $R$ (corresponding to density variance
$S_R$) and mean density fluctuation $\del$ is \citep{BLflucts} \beqa
\lefteqn{f_{\rm bias}(\del_c(z),\delta,S_R,S) = \frac{f_{\rm
ST}(\del_c(z),S)} {f_{\rm PS}(\del_c(z),S)} \ f_{\rm PS}
(\del_c(z)-\delta,S-S_R) \ ,} \nonumber \\ && \label{eq:bias} \eeqa
where $f_{\rm PS}$ and $f_{\rm ST}$ are, respectively, the
Press-Schechter and Sheth-Tormen halo mass functions. The value of
$F_{\rm coll}(\del_c(z),\delta,S_R,S)$ is the integral of $f_{\rm
bias}$ over $S$, from 0 up to the value $S_{\rm min}$ that corresponds
to the minimum halo mass $M_{\rm min}$ or circular velocity $V_{\rm
c}=\sqrt{G M_{\rm min}/R_{\rm vir}}$ (where $R_{\rm vir}$ is the
virial radius of a halo of mass $M_{\rm min}$ at $z$). We then
numerically find the value of $\delta$ that gives $\zeta F_{\rm
coll}=1$ at each $S_R$, yielding the exact barrier.  Also, in order to
compare with a simpler, analytically-solvable model, we derive a
linear approximation to the barrier, $\delta(S_R) \approx \nu + \mu
S_R$, by numerically finding the value of the barrier at $S_R=0$ and
its derivative with respect to $S_R$. In general, photon conservation
implies that the mean global ionized fraction should equal $\bar{x}^i
= \zeta \bar{f}_{\rm ST}$ in terms of the cosmic mean collapse
fraction.

\citet{b07} and \citet{diffPDF} used an approximation in which
effectively each factor on the right-hand side of
equation~(\ref{eq:bias}) was integrated separately over $S$, yielding
a simple analytical formula for the effective linear barrier. This
approximation was also assumed by \citet{fmh06} when they stated that
this hybrid prescription does not change the bubble size distribution
from the pure Press-Schechter case (for a fixed redshift, minimum halo
mass, and cosmic mean ionized fraction). Here we solve numerically for
the barrier using the exact formulas. We show that the previously-used
approximation is not too accurate, especially at the early stages of
reionization that are our focus in this paper.

\subsection{The statistics of a random walk with a barrier and 
discrete sources}

\label{s:full}

The standard approach presented above treats the random walks as
functions of a continuous variable $S_R$, and assumes a one-to-one
correspondence between the value of $\del$ and the ionized fraction
$x^i$ at each scale. The statistical distribution of first barrier
crossing, which physically corresponds to the bubble size
distribution, can be derived analytically for the approximate linear
barrier \citep{fzh04}, and for the exact barrier can be solved with
Monte Carlo simulations of random walks or by solving an integral
equation \citep{zhang}.

In reality, there are two additional physical constraints that are
neglected in the standard approach: the ionizing sources are discrete,
and the ionized fraction (for a given value of $\del$ in a region)
fluctuates due to Poisson fluctuations in the number of galaxies.  The
discreteness of ionizing sources means that the possible volume of
bubbles has a minimum value $\Vb$ corresponding to the bubble due to a
single galaxy hosted by a halo of mass $M_{\rm min}$. Also, the
expected ionized fraction $x^i$ given by the continuous model is
subject to Poisson fluctuations, as the actual ionized fraction
depends on the number of galaxies. Unlike the standard random walk
approach, in which the statistics of the walk depend only on the
barrier expressed as a function $\delta(S_R)$, Poisson fluctuations
introduce an explicit dependence on the mapping between $S_R$ and
scale $R$.

In order to include these discrete aspects in the bubble distribution,
we begin with the standard analytical approach, which considers the
statistics of spherical volumes of various sizes $R$, all about a
point $A$. Given a value $\del$ on a scale $R$ (with corresponding
variance $S_R$), we now treat the continuous ionized fraction $x^i$ of
the previous subsection only as an average expectation value. To find
its real distribution, we first calculate the mean expected value
$\langle j \rangle$ of the number of ionizing sources within the
sphere of radius $R$. This (non-integer) value can be calculated from
the integral of $f_{\rm bias}\, dS$ weighted by $1/M$ (which yields
halos weighted by number rather than mass); it depends on $z$, $S_{\rm
min}$, $\del$, and $S_R$. The actual value of $j$ is given by a
Poisson distribution with mean equal to $\langle j \rangle$. To find
the actual $x^i$, we find the mass of each of the $j$ halos according
to the halo mass distribution given by $f_{\rm bias}$; note that this
procedure does not involve a single, fixed mass distribution since
$f_{\rm bias}$ is a function of $\del$ and $S_R$.

The complicating factor in this procedure is that we cannot treat each
scale $R$ independently, since the ionizing sources are correlated
among the various volumes. This is the case first because the
densities $\del$ are correlated, and second because the Poisson
fluctuations are correlated, since each sphere contains all the
galaxies that lie within all smaller enclosed spheres. The correlation
of the densities is dealt with in the standard way reviewed above,
where small-scale power is added gradually as smaller spheres are
considered. This makes $\del(S_2)$ dependent on $\del(S_1)$ if $S_2 >
S_1$, forcing us to start on large scales $S_R=0$ and go to smaller
ones. However, the Poisson fluctuations are correlated in the other
direction, since a region $S_2$ contains a region $S_1$ if $S_2 <
S_1$.

The solution is a two-step Monte Carlo method: first, we generate the
random walk $\del(S_R)$, going from $S_R=0$ to its maximum value
(corresponding to the minimum bubble volume $\Vb$) in equal steps. At
each $S_R$ step, we find the mean expected number of galactic halos
$\langle j \rangle (S_R)$ and the mean expected total mass of these
halos, $\langle M_{\rm tot} \rangle (S_R)$. Note that the mean
expected ionized fraction is $\langle x^i \rangle (S_R) = \zeta
\langle M_{\rm tot} \rangle (S_R) / M(S_R)$, where $M(S_R)$ is the
total mass contained within the spherical volume of radius $R$. In the
second step, we generate the actual ionized fractions starting from
the smallest scale, $\Vb$, and working outwards. At $\Vb$, we generate
an instance of a Poisson distribution with mean $\langle j \rangle
(S_R)$, yielding an actual integer number $j$ of halos, for each of
which we find its mass from the appropriate distribution of halo
number versus mass, derived from $f_{\rm bias}$. Then, for each larger
scale (i.e., smaller $S_R$ value), the additional number of galaxies
from the last step is on average expected to be the difference
$\langle \Delta j \rangle= \langle j \rangle (S_1) - \langle j \rangle
(S_2)$, where $S_1 < S_2$ are two consecutive steps in $S_R$. We find
the actual difference $\Delta j$ from a Poisson distribution with a
mean of $\langle \Delta j \rangle$. However, while an actual number of
galaxies cannot be negative, sometimes the random walk in $\del$ gives
a value $\langle \Delta j \rangle <0$. In this case we assume that
$\Delta j=0$, since the number of galaxies already enclosed in a
smaller volume (corresponding to $S_2$) must also be found in the
larger, enclosing volume ($S_1$). We do not discard the negative value
of $\langle \Delta j \rangle$ but add it in the next step to the next
value of $\langle \Delta j \rangle$, continuing until we reach a
positive expected mean value on which we can operate a Poisson
distribution. In each step, we also keep track of the expected total
mass difference, $ \langle \Delta M \rangle = \langle M_{\rm tot}
\rangle (S_1) - \langle M_{\rm tot} \rangle (S_2)$. We slightly
modify\footnote{We scale the input $M$ value of the cumulative
distribution of halo mass $M$ (so that the total probability of having
$M \ge 0$ remains unity), with the scaling factor (typically close to
unity) chosen to yield the correct mean halo mass.} each distribution
$f_{\rm bias}$ that we use to generate the individual halo masses so
as to give the correct expected $ \langle \Delta M \rangle$. This
procedure ensures that on each scale we obtain the correct average
number of galaxies and correct average galaxy mass, both to high
accuracy. We note also that in each step in $S_R$, even if $\del$ at
the end of the step is below the barrier, there is a chance that the
random walk hit the barrier during the step. We estimate this
probability using a linear barrier approximation applied separately to
each step, and if the walk hit the barrier then we raise $\del$ at the
end of the step to the barrier. This procedure greatly accelerates the
convergence of the results as a function of the total number of steps
adopted in $S_R$.

\subsection{Summary of models}

\label{s:sum}

We summarize here the various models for the bubble size distribution
that we consider and compare below.

\begin{enumerate}

\item Model A: The correct distribution as given by our full model.
The bubble size distribution is calculated with our Monte Carlo method
with discrete sources and Poisson fluctuations, as detailed in
section~\ref{s:full}. We also keep track of how many sources are
contained in each generated bubble, which allows us to find the
cumulative volume fraction contained in bubbles with at least $N$
sources, as a function of $N$.

\item Model B: The exact, continuous barrier (without Poisson
fluctuations or discreteness). We calculate the non-linear barrier
$\del(S_R)$ numerically, as detailed in section~\ref{s:prev}. We then
derive the bubble size distribution with a Monte Carlo method that
generates random walks and tracks where they first cross the barrier.

\item Model C: A continuous linear barrier approximation. We calculate
a linear barrier approximation $\delta(S_R) \approx \nu + \mu S_R$
numerically, as detailed in section~\ref{s:prev}. We then derive the
bubble size distribution analytically as in \citet{fzh04}.

\item Model D: The previously-used continuous linear barrier
approximation. Here we apply the additional approximation mentioned at
the end of section~\ref{s:prev}, where we noted that it gives the same
bubble size distribution as in the linear barrier approximation of the
pure Press-Schechter (rather than Sheth-Tormen) model. In this case we
calculate analytically a linear barrier approximation $\delta(S_R)
\approx \nu + \mu S_R$ and then derive the resulting bubble size
distribution analytically as in \citet{fzh04}.

\item Model E: The pure stochastic Poisson model suggested by
\citet{FurOh}. This model, which neglects halo correlations and
assumes randomly placed, equal-intensity sources, yields an analytical
result \citep{FurOh} for the cumulative volume fraction contained in
bubbles with at least $N$ sources.

\end{enumerate}

Note that the minimum bubble scale is $V_{\rm bub}$ for models A and
E, and $V_{\rm bub}/\zeta$ (corresponding to the scale of the minimum
halo mass $M_{\rm min}$) for models B--D. Also note that we have
tested our barrier-crossing Monte-Carlo code by comparing it to the
analytical solution of a continuous linear barrier (Models C and
D). We have confirmed precise convergence, to within a relative error
below $1\%$ in the total ionization probability, i.e., the total
probability of crossing the barrier.

\section{Results}

We illustrate our results for a wide range of possible parameters for
either hydrogen reionization or helium (full) reionization. In the
latter case, $\zeta$ is simply interpreted as the overall efficiency
factor of producing helium-ionizing photons in halos. For hydrogen,
minimum halo masses $M_{\rm min}$ that are often considered are the
minimum mass for atomic cooling (corresponding to a circular velocity
$V_c\sim 16.5$ km/s, where $V_c=\sqrt{GM/R}$ in terms of the halo
virial mass and radius), or (at very high redshift) the minimum mass
for molecular hydrogen cooling ($V_c \sim 4.5$ km/s). Also, much
larger minimum masses are possible, $V_c \sim 35$ km/s due to
photoionization feedback (which should affect most of the universe by
the time reionization is well advanced), or even larger values if
internal supernova feedback strongly decreases the star formation
efficiency of low-mass halos at high redshift. For helium
reionization, assuming it occurs much later, photoionization feedback
affects the source halos from an early stage when the density of the
assembling matter is still low, resulting in a larger $V_c \sim 80$
km/s (i.e., of order the Jeans mass). Furthermore, if the observed
super-linear local relation between halo and black hole mass holds at
high redshift, then quasars are relatively much brighter in more
massive halos, increasing the typical halos of helium-ionizing sources
to $V_c \sim 200-300$ km/s. For a given $V_c$, the efficiency $\zeta$
can be chosen to give complete reionization\footnote{Note that our
simple model does not include the likely added importance of
recombinations towards the end of reionization. However, in this paper
we do not consider the late stages of reionization except as a
convenient fiducial mark for normalizing $\zeta$ through $\zr$.}
$\bar{x}^i = 1$ at various redshifts $\zr$. For a fixed $\zr$, a
larger $M_{\rm min}$ implies that rarer halos caused reionization,
resulting in larger Poisson fluctuations.

\subsection{Basic results and comparison with previous models}

\label{s:1perc}

\begin{figure}
\includegraphics[width=84mm]{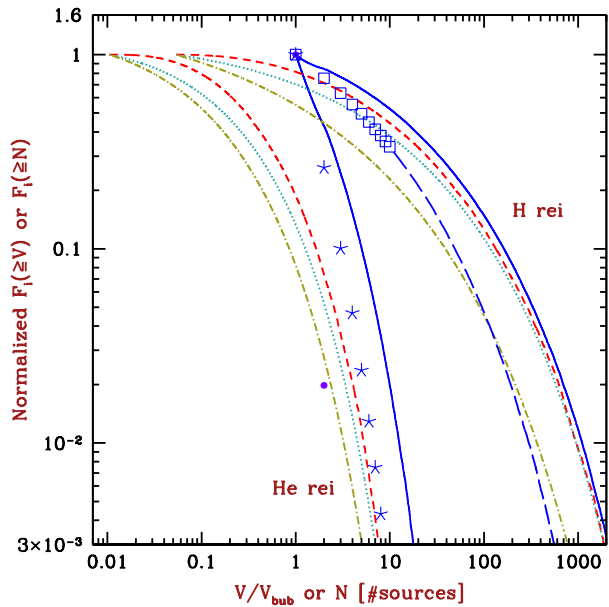}
\caption{Cumulative bubble volume distribution as a function of
$V/\Vb$, or of the number $N$ of ionizing sources in the
bubble. Assuming $\zr=7$ and $V_c= 16.5$ km/s for H, and $\zr=3$ and
$V_c = 285$ km/s for He, we consider $\bar{x}^i = 1\%$ ($z=16.3$ for
H, 6.3 for He). We compare $F_i(\ge V)$, the fraction of the ionized
volume contained in bubbles with volume $\ge V$, from our full Monte
Carlo method with discrete sources (Model A; solid curves) to $F_i(\ge
V)$ from a continuous barrier (Model B; short-dashed curves), a
continuous linear barrier approximation (Model C; dotted curves), and
a continuous linear Press-Schechter barrier (Model D; dot-dashed
curves). We also show $F_i(\ge N)$, the fraction of the ionized volume
contained in bubbles containing $\ge N$ sources, from our full model
(Model A; H: squares, He: stars; long-dashed curves for $N>10$), and
from a pure Poisson model (Model E; circles).}
\label{f:1perc}
\end{figure}

We begin by considering examples corresponding to an early stage (mean
ionized fraction $\bar{x}^i = 1\%$) of hydrogen or helium
reionization. For hydrogen, we assume atomic cooling ($V_c= 16.5$
km/s), with efficiency set to complete reionization (i.e., $\bar{x}^i
= 1$) at $\zr=7$ (implying $\zeta=19$). For helium we assume $\zr=3$
and $V_c = 285$ km/s (implying $\zeta=95$), which gives the bubbles a
minimum size at $z \sim 3$ of $R=10$ Mpc, about the expected size for
the quasars that are observed to dominate the ionizing photon
production at that redshift \citep{FurOh}. Figure~\ref{f:1perc} shows
that the bubble size distribution obtained from our full model is
substantially different from the predictions of previous models that
are based either on a continuous barrier or on a purely stochastic
Poisson distribution.

In general these models, based as they are on spherical statistics, do
not precisely conserve photons, and thus do not yield precisely the
desired $\bar{x}^i \equiv \zeta \bar{f}_{\rm ST}$ which we have set to
$1\%$. Indeed, the raw total ionization probability yielded by the
models is $2.1\%$ (H) and $2.0\%$ (He) for model D, $0.90\%$ (H) and
$1.4\%$ (He) for model C, and $1.3\%$ (H) and $1.4\%$ (He) for model
A. Thus, in the figure we compare the relative distributions,
expressed in terms of the fraction $F_i$ of the ionized volume
contained in various bubbles. Note that model E is defined according
to the desired $\bar{x}^i$, and model B (the exact continuous barrier)
is mathematically consistent in the sense that it yields the correct
total $\bar{x}^i$ if the probability is integrated down to $V=V_{\rm
bub}/\zeta$ (We have numerically verified this mathematical
consistency to a relative error of $\sim 1\%$).

Discreteness strongly fails for the continuous barrier models (both
linear and non-linear), in the sense that much of the ionized volume
in these models is predicted to occur inside bubbles below the minimum
volume $\Vb$, especially in the case of Helium reionization. Indeed,
$F_i(\ge \Vb)$ is only $55\%$ (H) and $8.2\%$ (He) for model D, $70\%$
(H) and $13\%$ (He) for model C, and $82\%$ (H) and $19\%$ (He) for
model B. Thus, the continuous barrier models fail since they assign a
substantial probability to the unphysical case of fractional bubbles
that are produced by less than one source. Expressed differently, the
continuous barrier models underpredict $F_i(\ge \Vb)$ since they do
not include the Poisson fluctuations that allow large regions to
sometimes reach $x^i=1$ even when their mean expected ionized fraction
$\langle x^i \rangle (S_R)$ is below unity.

Figure~\ref{f:1perc} also illustrates the continuous model with a
linearly approximated barrier, a model used very commonly because it
yields analytical predictions \citep{fzh04}. The error of the linear
barrier approximation grows at small scales, and becomes a $10\%$
error in the barrier height at $V \sim 0.07 \Vb$ (H) or $V\sim 0.02
\Vb$ (He). However, the linear barrier approximation becomes
relatively accurate on scales larger than the scale $\Vb$
corresponding to a one-source bubble. On that scale, the height of the
linear barrier in the examples shown here is only slightly below the
height of the real barrier (by $2.6\%$ for H and just $0.05\%$ for
He), though when $\bar{x}^i \ll 1$ the barrier corresponds to a rare
$\sim 3$--$\sigma$ fluctuation on this scale (and rarer still at
larger scales), and thus small differences in barrier height translate
to larger differences in $F_i$. The figure also shows that the pure
Press-Schechter model (model D) is a rather poor approximation to
model C. The Sheth-Tormen hybrid model yields more large bubbles than
the Press-Schechter model, which agrees with the expectation based on
the Sheth-Tormen mean halo mass function, which yields more rare,
massive halos than does the Press-Schechter mass function.

\begin{figure}
\includegraphics[width=84mm]{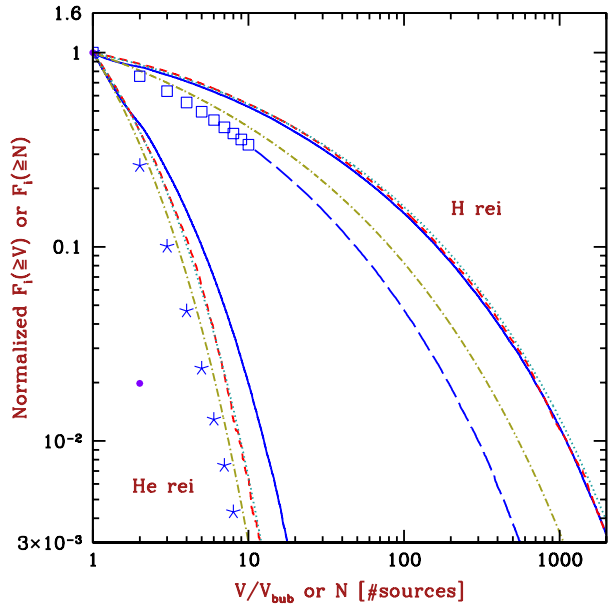}
\caption{Cumulative bubble volume distribution as a function of
$V/\Vb$, or of the number $N$ of ionizing sources in the bubble.  Same
as Figure~\ref{f:1perc}, except that the continuous models (models
B--D) have been cut off below $V = \Vb$ and renormalized at that
point. Note that the curves for models B and C nearly overlap.}
\label{f:1perc2}
\end{figure}

While the continuous barrier model extends unphysically to $V < \Vb$,
it does indicate correctly the fact that $F_i(\ge V)$ declines much
more rapidly with $V$ for the He case we consider than for H
reionization. In fact, we find that if we simply cut off the $V < \Vb$
portion and renormalize the continuous models relative to $V = \Vb$
(which is not a standard way of interpreting these models), then the
exact and linear barrier models yield nearly identical results, and
they both yield a reasonable rough estimate to the true bubble size
distribution in the full model. This is illustrated in
Figure~\ref{f:1perc2}, which shows the same quantities as in
Figure~\ref{f:1perc} except that all the continuous models have been
renormalized and are plotted only for $V \ge \Vb$. For instance, the
ratio $V_{1/2} \equiv F_i(V\ge \Vb)/F_i(V\ge 2 \Vb)$ is 1.18 (H) and
2.33 (He) in the full model (model A), 1.14 (H) and 2.54 (He) for the
continuous exact barrier (model B), and 1.15 (H) and 2.58 (He) for the
continuous linear barrier (model C). This approach to the continuous
models provides a reasonable estimate of the full bubble size
distribution in the case of H reionization; e.g., $V_{1/100} \equiv
F_i(V\ge \Vb)/F_i(V\ge 100 \Vb)$ for H is 6.73 in model A, 6.46 in
model B, and 6.21 in model C, so that here the linear model C,
calculated analytically (i.e., without using Monte Carlo random walks
or Poisson fluctuations), yields an estimate of $V_{1/100}$ that is
within $8\%$ of the true answer according to model A. However, for He
reionization this approach is much less successful in predicting
ratios involving large volumes; e.g., $V_{1/5} \equiv F_i(V\ge
\Vb)/F_i(V\ge 5 \Vb)$ for He is 9.9 in model A, 16.0 in model B, and
17.6 in model C, and these differences increase with $V$
(Figure~\ref{f:1perc2}).

With our full model (model A), we can also separately predict the
distribution by number $F_i(\ge N)$. This drops more rapidly with $N$
than the distribution by volume $F_i(\ge V)$ does with $V$, since
large-volume bubbles can be produced either by having many sources of
mass $\sim M_{\rm min}$ or with a smaller number of individually
massive halos taken from the high-mass end of the halo mass
function. Still, $F_i(\ge N)$ declines with $N$ much more slowly than
a pure Poisson model would predict. Indeed, a purely stochastic model
as suggested by \citet{FurOh} for the early stages of He ionization
(or even as late as $\bar{x}^i \sim 50\%$), where Poisson fluctuations
are assumed that are uncorrelated with the underlying density
distribution, completely fails to describe the results. The analytical
predictions of this model \citep{FurOh} yield, for $\bar{x}^i =1\%$
(for either H or He), $F_i(N\ge 2)=2.0\times 10^{-2}$ and $F_i(N\ge
3)=4.4\times 10^{-4}$ (with the latter already outside the range of
Figures~\ref{f:1perc} and \ref{f:1perc2}). In particular, the ratio
from the previous paragraph (but applied to the number of sources),
$N_{1/2} \equiv F_i(N\ge 1)/F_i(N\ge 2)$, is 1.32 (H) or 3.8 (He) in
the full model, compared to $N_{1/2} = 51$ for model E. Clearly,
density correlations play a substantial role in determining the
abundance of multi-source bubbles, even early on in reionization and
even when the process is driven by large, rare ionizing sources (such
as quasars).

\begin{figure}
\includegraphics[width=84mm]{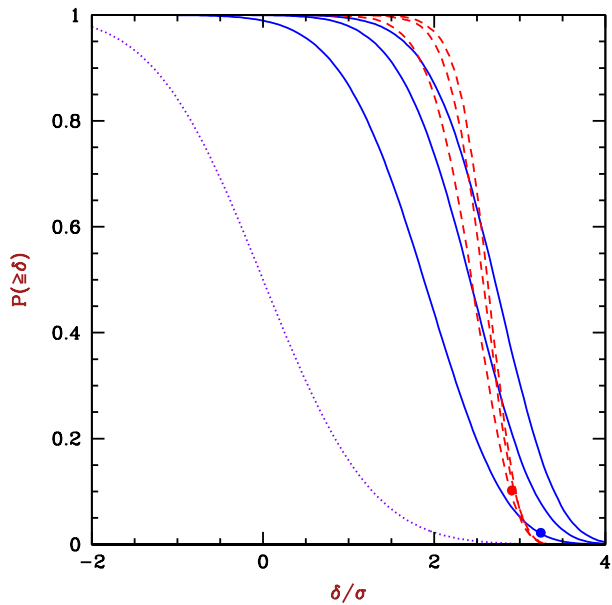}
\caption{Cumulative probability distribution of the linear density
$\del$ in units of its standard deviation $\sigma=\sqrt{S}$ on the
relevant scale. For either H (dashed curves) or He (solid curves)
reionization with the same parameters as in Figure~\ref{f:1perc}, we
show $P(\ge \del)$ for regions ionized by exactly one, two, or three
sources (from left to right in each set of curves). In each case, a
circle on the one-bubble curve shows the median barrier height on the
corresponding scale. Also shown for comparison is the cumulative
distribution of the normal distribution for unconstrained regions
(dotted curve).}
\label{f:del}
\end{figure}

To help understand the relation of the full model to the pure Poisson
and to the continuous barrier models, we show in Figure~\ref{f:del}
the relation between ionization in bubbles and the underlying linear
density $\del$. Density fluctuations are strongly correlated with
ionization, so that the density of ionized regions is strongly biased
high, and the distribution is very different from the standard
Gaussian that would be expected in a pure Poisson model. However,
Poisson fluctuations allow regions to fully ionize themselves even if
their density is significantly lower than the barrier, which in a
continuous model would set the minimum needed $\del$ for ionization by
internal sources. In particular, the median $\del$ for regions ionized
by exactly $N$ sources (where 'exactly' means not contained in any
larger H~II region) represents a fluctuation of 2.4-$\sigma$,
2.57-$\sigma$, and 2.61-$\sigma$ (for H) or 1.9-$\sigma$,
2.4-$\sigma$, and 2.7-$\sigma$ (for He), for $N=1$, 2, and 3,
respectively. The corresponding (median) barriers, on the other hand,
are 2.907-$\sigma$, 2.909-$\sigma$, and 2.922-$\sigma$ (for H), or
3.2-$\sigma$, 3.5-$\sigma$, and 3.7-$\sigma$ (for He). Thus, the
barriers do give a good rough indication in each case of whether the
$\del$ distributions for various $N$ are spaced out or squeezed
together. This in turn determines whether one-source bubbles are
dominant and $N>1$ is rare, or if multi-source bubbles are at least as
common as $N=1$.

The continuous model indicates that the main parameters controlling
the relative dominance of single-source bubbles are the effective
efficiency $\zeta$ and the effective slope of the power spectrum on
the scale $R_{\rm bub}$ of a one-source bubble. The efficiency sets
the ratio between the scale $R_{\rm bub}$ of a one-source bubble and
the scale $R_{\rm min}$ from which a galactic halo of mass $M_{\rm
min}$ was assembled (this ratio equals $\zeta^{1/3}$). Now, the key
issue is the relative difficulty of each scale achieving
self-ionization, when we consider different scales. To self-ionize, a
region must reach a high enough collapse fraction, which according to
the extended Press-Schechter formula in equation~(\ref{eq:exPS}),
requires a value of $\delta$ that depends on the variance $(S_{\rm
min} - S_R)$ available for density fluctuations inside the
region\footnote{The Sheth-Tormen hybrid model alters things slightly,
but we use the simpler formula here as a rough guide for a qualitative
understanding.}. In order to reach this required value (i.e., the
barrier), the density has the variance $S_R$ to work with. Thus, when we
increase the scale (e.g., going from a typical one-source bubble to
one with two sources), if the fractional decline in $S_R$ is more rapid
than in $(S_{\rm min} - S_R)$, then self-ionized regions become rarer
quickly with increasing scale, leading to the dominance of one-source
bubbles. This is the case when $S_R \ll S_{\rm min}$, i.e., it requires
first that the bubble and halo scales differ by a large factor (which
requires large values of $\zeta$), and also that the variance depend
significantly on scale (otherwise, $S_R$ and $S_{\rm min}$ will be about
the same even if the corresponding scales are very different).

More quantitatively, the fractional decline in $S_R$, over the
fractional decline in $(S_{\rm min} - S_R)$, is (for small changes in
$S_R$) equal to $(S_{\rm min}/S_R - 1)$. If the power spectrum of density
fluctuations is approximated as a power law with an effective index
$n$ over the relevant range of scales, then this ratio, which
indicates how much harder (in terms of number of $\sigma$ of the
fluctuation) it is to ionize larger scales, is approximately
$\zeta^{1+(n/3)}-1$. On small scales, $n$ approaches the asymptotic
value of $-3$, making all scales behave roughly equally even when
$\zeta$ is relatively large. Note, though, that increasing $\zeta$
increases $R_{\rm bub}$ and thus brings larger scales into play,
making the effective $n$ less negative and thus boosting the effect of
the increased $\zeta$ on making few-source bubbles dominant. This puts
a quantitative face on the intuition that rare sources tend to create
bubbles with small numbers of sources. To illustrate, in our H example
$R_{\rm min} = 64$ kpc and $R_{\rm bub} = 169$ kpc, giving $n \sim
-2.5$, while in the He example $R_{\rm min} = 1.7$ Mpc and $R_{\rm
bub} = 7.8$ Mpc, giving $n \sim -2$. Thus, He reionization by quasars
has both a high efficiency and corresponds to a relatively large
scale, both of which contribute to making small bubbles more dominant,
in particular the smallest bubbles created by single sources.

\subsection{Approximate calculation}

\label{s:aprx}

The current lack of observations at high redshifts leaves basic
parameters of the galaxy population unconstrained at early
times. While our model can be used to calculate the bubble
distribution in any particular case, the need to run Monte Carlo
trials makes it difficult to explore a large parameter space. Thus, an
approximate but quick calculation is useful for this purpose. In
developing such an approximation, we focus on determining when
one-source bubbles dominate the ionizing volume. This can be
investigated with the ratio $N_{1/2}$, which is close to unity when
multi-source bubbles dominate, and is $\gg 1$ when one-source bubbles
do. Specifically, this ratio is related to the fraction $F_i(N=1)$ of
the ionized volume that is contained in one-source bubbles through
$N_{1/2} = 1/[1-F_i(N=1)]$.

To construct an approximate calculation of this ratio we first adopt
the approximation of having equal intensity sources, all corresponding
to halos having a mass equal to the mean expected mass $\langle M
\rangle$. While this approximation does not work well for obtaining
information on the bubble size distribution, we find that it works
reasonably for our desired ratio involving the distribution of number
of sources per bubble. We first consider in general the
self-ionization probability on the scale of a bubble containing $j$
sources (with a variance $S$ that we approximate as that corresponding
to a volume $j \Vb$), i.e., the probability that a region of this size
contains at least $j$ sources (regardless of whether or not it is
contained in some larger bubble). A first attempt to calculate this
quantity $P_{\rm self}(j)$ is to calculate the Poisson probability of
having at least $j$ sources, averaged over the normal distribution of
$\del$ on the scale $S$: \beq P_{\rm self}(j) = \int d\del \frac{1}
{\sqrt{2 \pi S}} e^{-\del^2/(2 S)} P_{\rm Pois}(j \tilde{x}^i(\del,S);
\ge j)\ ,
\label{eq:self} \eeq where $P_{\rm Pois}(\alpha;\ge j)$ denotes the
probability of having at least $j$ sources in a Poisson distribution
with mean $\alpha$, and $\tilde{x}^i$ (which also depends on $z$ and
$S_{\rm min}$) is an approximate estimate of $\langle x^i \rangle$
where we use the same approximation as in model D in order to obtain a
simple formula. For large bubbles, equation~\ref{eq:self} for $P_{\rm
self}(j)$ underestimates the self-ionization probability, since for a
given mean $\del$ in the region, internal density fluctuations
increase the variance of the number of sources beyond a pure Poisson
distribution. For $j=2$ we can instead calculate a more accurate
self-ionization probability by calculating a double integral over the
joint normal distribution of $\del_1$ and $\del_2$, the mean densities
inside a one-source volume $\Vb$ and inside the surrounding two-source
volume, respectively. Given $\del_1$ and $\del_2$, the mean expected
number of sources in the two regions is $n_1=\tilde{x}^i(z,S_{\rm
min},\del_1,S_1)$ and $n_2=2 \tilde{x}^i(z,S_{\rm min},\del_2,S_2)$,
respectively, where $S_1$ and $S_2$ are the corresponding
variances. The probability of self-ionization of the two-source volume
is then the probability of having at least 2 total sources from the
sum of a Poisson distribution of mean $n_1$ plus a Poisson
distribution of mean $n_2-n_1$ (except that the latter quantity is
restricted to be non-negative, a key point which allows the larger
fluctuations in $n_1$ to contribute).

Calculating $P_{\rm self}(j)$ exactly for $j>2$ would require at least
a triple integration, but since $j=1$ and $j=2$ are most important for
estimating $N_{1/2}$, we simply estimate the self-ionization
probability for all $j>2$ with equation~(\ref{eq:self}). Now, $P_{\rm
self}(j)$ for any $j$ is itself only a lower limit for the ionization
probability $P(N\ge j)$, since the region may be part of a larger H~II
bubble even if it cannot fully ionize on its own. Actually, when
one-source bubbles dominate and $P(N\ge j)$ drops rapidly with $j$,
regions are much more likely to self-ionize than to get ionization
help from larger scales, and then $P_{\rm self}(j)$ becomes an
accurate estimate of $P(N\ge j)$. However, in order to achieve
reasonable accuracy also when multi-source bubbles are important, we
add a correction to each $P_{\rm self}(j)$ based on the values of
$P_{\rm self}(k)$ for $k>j$. Indeed, instead of just calculating
$P_{\rm self}(k)$, which is the probability of having at least $k$
sources in a region of size corresponding to $k$ sources, we can
separately estimate $P_{l}(k)$, the probability of having exactly $l$
sources in that region, using a formula just like
equation~(\ref{eq:self}) but using the Poisson probability of finding
$l$ sources. Then, for any number $l\ge k$ sources, we calculate the
additional ionization probability that was not previously included in
$P(N\ge j)$ (for each internal volume $j<k$) using the approximation
that the $l$ sources are uniformly distributed within the volume
$k$. In this way, we estimate the probabilities $P(N\ge 1)$ and
$P(N\ge 2)$ including the contributions of larger volumes with
$j>2$. When one-source bubbles dominate, higher-$j$ volumes have a
small effect, but when multi-source bubbles dominate the effect adds
up, and we cut off $j$ so that $P(N\ge 1)$ does not rise above the
global ionization fraction $\bar{x}^i $. Actually, we find that while
the correction from higher-$j$ volumes can change each of $P(N\ge 1)$
and $P(N\ge 2)$ by up to a factor of a few (giving results much closer
to the full model A), the relative effect on their ratio is $\sim
15\%$ at most.

Our estimate for $N_{1/2}$ is simply $P(N\ge 1)/P(N\ge 2)$. The
approximate calculation becomes exact in the limit $N_{1/2}
\rightarrow \infty$, where our estimated probabilities $P_{\rm
self}(j)$ become very small for all $j \ge 2$, while in the opposite
limit, when $N_{1/2} \rightarrow 1$ all quantities become nearly
independent of $j$ and thus our estimate for the ratio approaches
unity, also correctly. In practice, from direct comparison with the
Monte Carlo method at $\bar{x}^i$ ranging from $10^{-6}$ to 1, and at
ratios $N_{1/2}$ ranging from 1 to 200, we find that our approximation
for this ratio is accurate to $\sim 15\%$ (though below we extrapolate
it beyond the tested range).
 
\begin{figure}
\includegraphics[width=84mm]{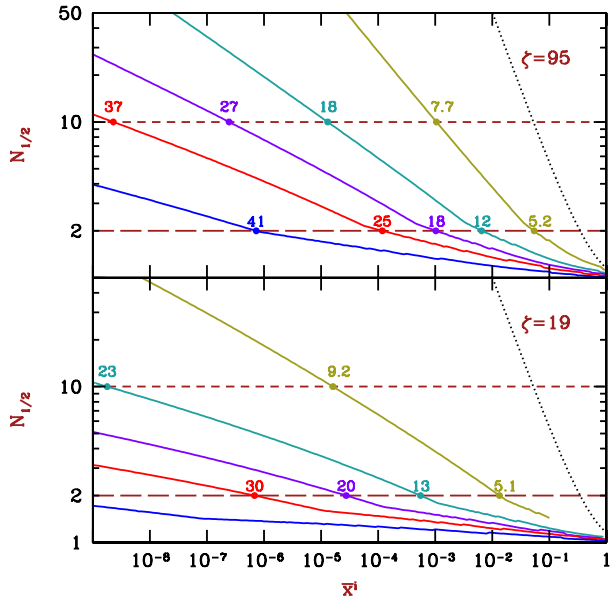}
\caption{Sweep of the parameter space using our approximate
calculation, showing the relative dominance of one-source compared to
many-source bubbles as indicated by the ratio $N_{1/2}=P(N\ge
1)/P(N\ge 2)$. For $\zeta=19$ or 95, as indicated, we consider
galactic halos with minimum $V_c = 4.5$, 16.5, 35, 80, or 285 km/s
(solid curves, from bottom to top). We compare to the case of a pure
stochastic Poisson distribution (model E; dotted curves). Also shown
are the locations corresponding to half of the volume being in
one-source bubbles (horizontal long-dashed line), and to $90\%$ in
one-source bubbles (horizontal short-dashed line); redshifts are
indicated at these locations for each case (if it lies within the
range of the plot). Note also that the various curves are not
continued below $z=3$ (for $V_c = 285$ km/s) or $z=6$ (for the other
cases).}
\label{f:scan1}
\end{figure}

\begin{figure}
\includegraphics[width=84mm]{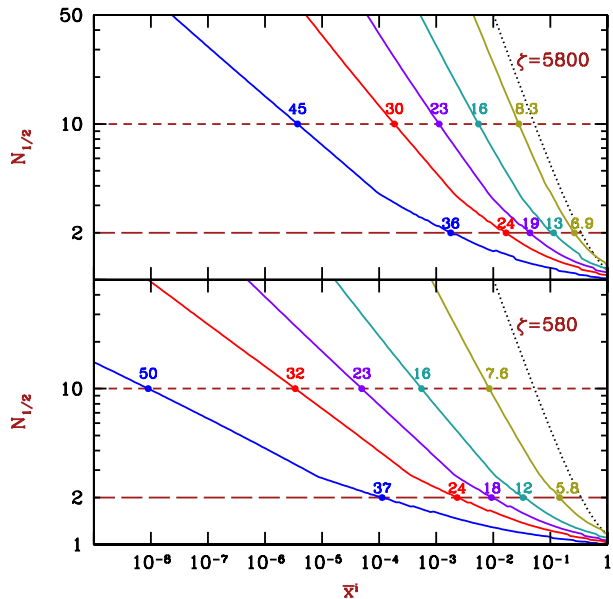}
\caption{Same as Figure~\ref{f:scan1} but for $\zeta=580$ or 5800, as
indicated.}
\label{f:scan2}
\end{figure}

Having developed a quick, relatively accurate calculation method, we
can use it to explore which areas of parameter space will be dominated
by one-source bubbles and which will form many multi-source
bubbles. Figures~\ref{f:scan1} and \ref{f:scan2} show the ratio
$N_{1/2}$ in the approximate calculation, for $\bar{x}^i$ ranging from
1 down to $10^{-9}$, over the whole relevant range of source masses,
i.e., assuming a minimum $V_c = 4.5$, 16.5, 35, 80, or 285 km/s, and
for four values of the efficiency $\zeta$, 19, 95, 580, and 5800. It
is interesting to consider the whole parameter space, without
normalizing to a particular reionization redshift, since the dominant
population of ionizing sources at any given redshift may not have
similar properties to that near the end of reionization, due to the
evolution in time of chemical, radiative, and hydrodynamical
feedbacks.

The above values of $\zeta$ are chosen to be particular interesting,
where $\zeta=19$ corresponds to $\zr=7$ for $V_c= 16.5$ km/s, and
$\zeta=95$ corresponds to $\zr=3$ for $V_c = 285$ km/s, which are the
H and He reionization examples considered in the previous
subsection. More generally for star-forming halos, Population II stars
(assumed similar to low-metallicity stars forming today) produce $\sim
5800$ ionizing photons per hydrogen atom in stars, while Population
III stars (assumed to consist of $100 M_\odot$, zero-metallicity
stars) produce around 10 times more. Thus, if we assume a maximum star
formation efficiency of $10\%$ (i.e., that this fraction of the
baryons in a halo are contained in stars), then if all ionizing
photons escape out of the dense surroundings of the stars and the
halo, we get a maximum possible $\zeta=5800$, with Pop III stars. A
value of $\zeta=580$ can then represent several possibilities: perhaps
only $10\%$ of photons escape, or $100\%$ escape but we assume Pop II
stars, or we assume Pop III stars but with a star formation efficiency
of only $1\%$. The latter value is indeed the efficiency expected for
the very first, primordial Pop III stars in molecular cooling
halos. Numerical simulations suggest that in each $\sim 10^5 M_\odot$
halo (containing, therefore, $\sim 10^4 M_\odot$ in baryons), it is
likely that only a single $100 M_\odot$ star forms \citep{yoshida}
before its feedback disrupts the rest of the halo gas and prevents the
formation of additional stars, at least for some time. Note also that
these values of $\zeta$ neglect recombinations, which can only lower
the effective $\zeta$ further.

Figures~\ref{f:scan1} and \ref{f:scan2} imply the general conclusion
that ionizing sources produce isolated, single-source bubbles only
quite early in reionization, when $\bar{x}^i \ll 1$. This is a result
of the fact that while Poisson fluctuations are large when we consider
just one or two sources, they are strongly modulated by halo bias due
to the underlying density fluctuations. Thus, sources are usually
found in high-density regions, which makes it relatively likely to
find other sources nearby. As sources become rarer at high redshift,
the increasing correlation strength between halos partially
compensates for the overall low number density of sources, though
eventually the sheer rarity of sources does come to dominate. As
discussed above, increasing $V_c$ or $\zeta$ at a given $\bar{x}^i$
makes sources rarer and brings larger scales into play, making it
easier to form one-source bubbles relative to multi-source
bubbles. However, only the most extreme case we consider of rare,
extremely bright sources ($V_c = 285$ km/s and $\zeta=5800$, an highly
unlikely combination) approaches the results expected for a pure
stochastic Poisson distribution; the ratio in the stochastic model is
$N_{1/2}=1/[1-\exp(-2 \bar{x}^i)]$ \citep{FurOh}.

The Figures also indicate the redshifts when the fraction $F_i(N=1)$
of the ionized volume that is contained in one-source bubbles equals
$50\%$ (corresponding to $N_{1/2} =2$) or $90\%$ ($N_{1/2} =10$). In
particular, for $V_c= 16.5$ km/s normalized to produce H reionization
at $\zr=7$ (i.e., $\zeta=19$), one-source bubbles dominate (i.e.,
$F_i(N=1) > 90\%$) only above $z=57$ (outside the plot range), while
multi-source bubbles become equally important (i.e., $F_i(N=1)=50\%$)
at redshift 30. Primordial Pop III stars with $\zeta=580$ and $V_c=
4.5$ km/s also tend to form multi-source bubbles at rather high
redshifts, with one-source bubbles remaining dominant only down to
$z=50$, and with multi-source bubbles becoming equally important at
$z=37$. On the other hand, for He reionization at $\zr=3$ with $V_c=
285$ km/s (i.e., $\zeta=95$), these milestones are reached at $z=7.7$
and $z=5.2$, respectively. Additional cases where these milestones
occur outside the plot range of the Figures include $\zeta=95$ and
$V_c = 4.5$ km/s, which reaches $N_{1/2}=10$ at $z=62$; $\zeta=19$ and
$V_c = 35$ km/s, which reaches $N_{1/2}=10$ at $z=37$; and the
faintest example we consider for individual sources, $\zeta=19$ and
$V_c = 4.5$ km/s, which reaches $N_{1/2}=2$ at $z=53$ and does not
reach $N_{1/2}=10$ even at the most likely redshift ($z=65$) of the
very first star \citep{1st}.

If we consider a range of values of $\zeta$ for halos of a given
$V_c$, the global ionized fraction $\bar{x}^i$ corresponding to a
particular milestone (as defined by a particular value of $F_i(N=1)$)
increases with $\zeta$, since increasing $\zeta$ at a fixed
$\bar{x}^i$ makes sources rarer, while increasing $\bar{x^i}$ (with a
fixed $\zeta$) compensates for this by increasing the source number
density. For each milestone, however, the redshift, which
observationally is the most directly relevant quantity, behaves in a
more complicated way, since it is directly related to the number
density of sources, and thus depends on the ratio $\bar{x}^i / \zeta$.
We find that sources with a given $V_c$ can only achieve a dominance
of one-source bubbles at high redshift, almost regardless of the
efficiency $\zeta$ (and thus, regardless of the reionization
redshift). 

\begin{figure}
\includegraphics[width=84mm]{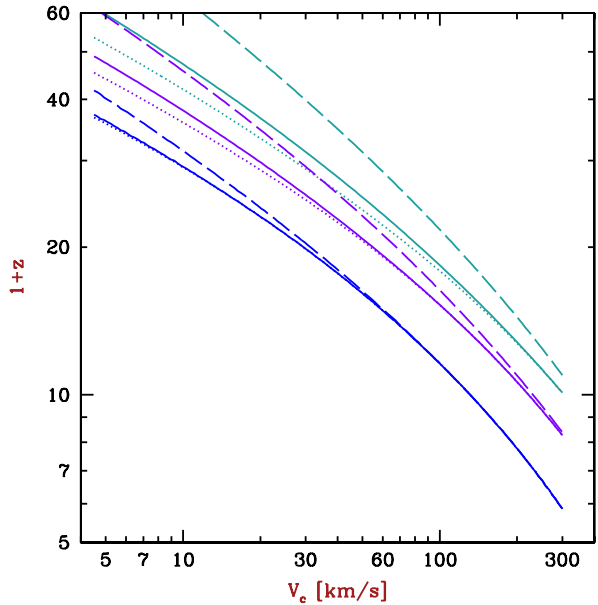}
\caption{Minimum redshift (shown in terms of $1+z$) required to
achieve a dominance of one-source bubbles, for ionizing sources in
halos with a minimum circular velocity $V_c$ (shown over the range
$4.5-300$ km/s). The minimum redshift shown here is required
regardless of the reionization redshift or the ionizing efficiency, as
long as $\zeta$ is in the range 10--100 (dashed curves), 10--1000
(solid curves), or 10--10000 (dotted curves). We consider milestones
when the volume fraction in one-source bubbles is $50\%$, $90\%$, or
$99\%$ (from bottom to top in each set of curves).}
\label{f:z}
\end{figure}

Figure~\ref{f:z} shows the minimum $z$ required to achieve various
values of $F_i(N=1)$ (as a function of $V_c$), assuming only that the
value of $\zeta$ lies within some wide range. The figure shows that
while high values of $\zeta$ do have a larger effect on low-$V_c$
halos, the minimum redshift is overall relatively insensitive to the
particular range assumed. In particular, assuming $10<\zeta<1000$, for
He reionization by quasars (assuming $V_c \le 300$ km/s), the volume
fraction in one-source bubbles $F_i(N=1)$ can be greater than $50\%$
only at $z>4.9$, $90\%$ at $z>7.3$, and $99\%$ at $z>9.1$. For H
reionization by stars (assuming $V_c \le 35$ km/s), these milestones
require $z>18$, $z>23$, and $z>28$, respectively. The generation of
atomic-cooling halos ($V_c = 16.5$ km/s) can achieve $F_i(N=1) > 50\%$
only at $z>24$, $90\%$ at $z>31$, and $99\%$ at $z>38$. Finally (again
assuming $10<\zeta<1000$), the earliest generation of
molecular-hydrogen-cooling halos ($V_c = 4.5$ km/s) can achieve these
milestones only at $z>36$, $z>48$, and $z>61$, respectively.

\subsection{Later stages}

\label{s:late}

\begin{figure}
\includegraphics[width=84mm]{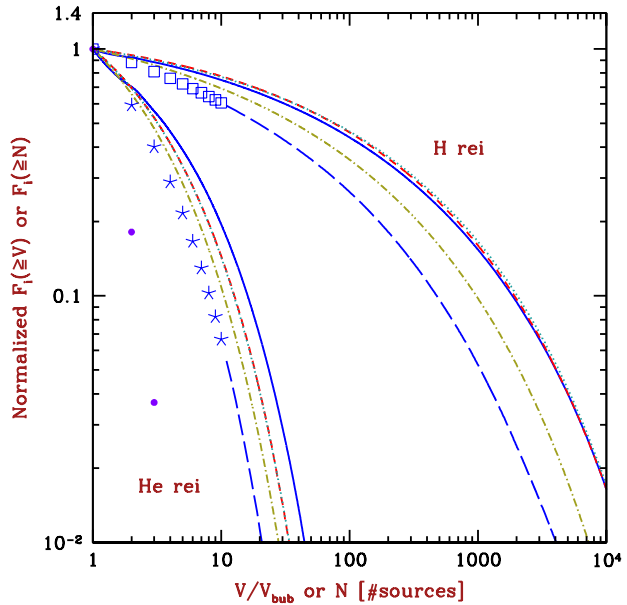}
\caption{Cumulative bubble size distribution as a function of $V/\Vb$,
or of the number $N$ of ionizing sources in the bubble. Assuming
$\zr=7$ and $V_c= 16.5$ km/s for H, and $\zr=3$ and $V_c = 285$ km/s
for He, we consider $\bar{x}^i = 10\%$ ($z=12.1$ for H, 4.8 for
He). We compare $F_i(\ge V)$, the fraction of the ionized volume
contained in bubbles with volume $\ge V$, between model A (solid
curves), model B (short-dashed curves), model C (dotted curves), and
model D (dot-dashed curves). We also show $F_i(\ge N)$, the fraction
of the ionized volume contained in bubbles containing $\ge N$ sources,
from model A (H: squares, He: stars; long-dashed curves for $N>10$)
and model E (circles). Note that the curves for models B and C
essentially overlap.}
\label{f:late1}
\end{figure}

\begin{figure}
\includegraphics[width=84mm]{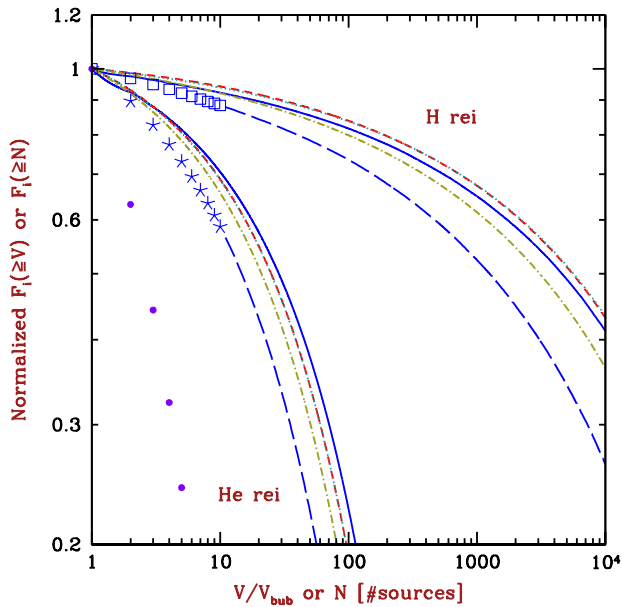}
\caption{Cumulative bubble size distribution as a function of $V/\Vb$,
or of the number $N$ of ionizing sources in the bubble. Same as
Figure~\ref{f:late1}, except calculated when $\bar{x}^i = 50\%$
($z=8.7$ for H, 3.6 for He). Note that the curves for models B and C
essentially overlap.}
\label{f:late2}
\end{figure}

As reionization advances, eventually the typical bubble size
encompasses a large number of ionizing sources, reducing the
importance of discreteness and of Poisson
fluctuations. Figures~\ref{f:late1} and \ref{f:late2} show the
cumulative bubble size distribution as in Figure~\ref{f:1perc2}, but
for later stages of reionization. At these times, the continuous
barrier models still have a significant probability at $V <\Vb$,
especially for He reionization by quasars; however, if only the $V >
\Vb$ portion is considered as in these figures (see also the
discussion in section~\ref{s:1perc}), then the linear barrier
predictions become essentially identical to those of the exact
barrier, and the predicted bubble size distributions of these
continuous models are reasonably accurate. Specifically, when
$\bar{x}^i = 10\%$, for the example of H reionization, $V_{1/2}$ and
$V_{1/100}$ equal 1.08 and 2.28, respectively, in model A (the full
model), 1.06 and 2.18 in model B (continuous barrier), and 1.06 and
2.16 in model C (linear barrier). For He reionization, $V_{1/2}$ and
$V_{1/5}$ are 1.42 and 2.56 in model A, 1.44 and 2.98 in model B, and
1.44 and 2.99 in model C. When the universe is $10\%$ ionized, bubbles
with a small number of sources still play a major role, e.g., one and
two-source bubbles together account for $19\%$ (H) or $60\%$ (He) of
the total ionized volume, and the small-$N$ regime is still quite
important.

At the midpoint of global reionization ($\bar{x}^i = 50\%$), the
continuous barrier models approach the full model even more (note that
the figures at different $\bar{x}^i$ have different $y$-axis
ranges). For the example of H reionization, $V_{1/2}$ and $V_{1/100}$
equal 1.03 and 1.22, respectively, in model A (the full model), 1.01
and 1.19 in model B (continuous barrier), and 1.01 and 1.19 in model C
(linear barrier). For He reionization, $V_{1/2}$ and $V_{1/5}$ are
1.08 and 1.23 in model A, 1.08 and 1.24 in model B, and 1.08 and 1.24
in model C. For H reionization only $5.2\%$ of the ionized volume lies
in one and two-source bubbles, but for He this fraction is still
$17\%$. As we found in section~\ref{s:1perc} at $\bar{x}^i = 1\%$, at
$\bar{x}^i = 10\%$ and $50\%$ we again see that the pure
Press-Schechter model (Model D) is a rather poor approximation to
model C, and that the pure Poisson model (Model E) predicts a
distribution by number $F_i(\ge N)$ that falls off much faster with
$N$ than do the true distributions (for H or He reionization)
according to model A.

\section{Conclusions}

\label{s:conc}

We have developed a model of reionization that adds discrete ionizing
sources and Poisson fluctuations to the continuous model of
\citet{fzh04}. We have shown how to obtain the distribution of ionized
bubbles, versus both bubble size and number of ionizing sources, with
a two-step Monte Carlo method that accounts for both density and
Poisson correlations among regions of various sizes surrounding a
given random point in the universe. The bubble size distribution we
obtained differs substantially from previous models, but if the
continuous barrier model is cut off below $\Vb$ (the minimum bubble
volume corresponding to a single halo of mass $M_{\rm min}$) then it
yields a reasonable rough estimate to the true bubble size
distribution. More specifically, this estimate is generally accurate
for H reionization even as early as a mean ionized fraction $\bar{x}^i
= 1\%$, while for He reionization it works best for small volumes and
at later times, and at $\bar{x}^i = 1\%$ is accurate only up to $V
\sim 3 \Vb$. Note that with the cutoff at $\Vb$, the linear barrier
approximation (which can be calculated analytically) gives nearly
identical results to the exact continuous barrier.

Our full model yields a bubble distribution by number $N$ that drops
more rapidly with $N$ than does the volume distribution drop with $V$,
but still, multi-source bubbles are always far more abundant than a
pure stochastic Poisson model would suggest. This is due to the fact
that density fluctuations are strongly correlated with ionization even
when Poisson fluctuations are large. Thus, the density of ionized
regions is strongly biased high compared to unconstrained regions, but
on the other hand, Poisson fluctuations allow regions to fully ionize
themselves even if their density is not as high as would be needed in
the continuous barrier model.

The main parameters controlling the relative dominance of
single-source bubbles are the effective efficiency $\zeta$ and the
effective slope $n$ of the power spectrum on the scale of a one-source
bubble. The ratio of how much harder (in terms of number of $\sigma$
of the fluctuation) it is to ionize large bubbles compared to small
ones, is approximately proportional to
$\zeta^{1+(n/3)}-1$. Reionization by rare sources that are massive and
bright corresponds to having a high $\zeta$ and to a high minimum
bubble size, which brings larger scales into play, making the
effective $n$ less negative and thus making it harder to produce
multi-source bubbles.

We have developed a quick, $15\%$ accuracy approximate calculation of
the ratio $N_{1/2}$ between the total ionized volume and that in
multi-source bubbles. This allowed us to sweep through the full
parameter space of possible halo masses and efficiencies of the
ionizing sources, and to show that sources with a given minimum
circular velocity $V_c$ can only achieve a dominance of one-source
bubbles at high redshift, regardless of their efficiency or of the
reionization redshift. In particular, for He reionization by quasars,
one-source bubbles can dominate (i.e., contain $90\%$ of the ionized
volume) only at $z>7.3$, and fill half the ionized volume at $z>4.9$,
while H reionization by stars can achieve these milestones only at
$z>23$ and $z>18$, respectively (assuming $10 < \zeta < 1000$). The
generation of atomic-cooling halos can place $90\%$ of the ionized
volume in isolated bubbles only at $z>31$ and $50\%$ at $z>24$, while
the earliest generation of molecular-hydrogen-cooling halos can
achieve the same only at $z>48$ and $z>36$, respectively.

We note that reality likely includes even more fluctuations than
included in our Poisson model, since we have still assumed that the
number of ionizing photons emitted from a galactic halo is
proportional to its mass. In reality, variations in the ionizing
efficiency (through spatial or temporal fluctuations in the star
formation efficiency and in the escape fraction of ionizing photons),
and in the merger histories of halos of a given mass (even within a
given environment, as measured by the average density of a surrounding
region) will increase the role of (now generalized) Poisson
fluctuations compared to that of galaxy bias due to the underlying
large-scale density fluctuations. Simple forms of such variability can
be included in a model of the type that we presented, since the
ionizing photon outputs from sources are added as individual units
(which could be generated from additional distributions for a given
halo mass). In general, the model we developed can be used to
investigate helium reionization and observational prospects for 21-cm
observations during the infancy of hydrogen reionization.

\section*{Acknowledgments}
The author thanks Andrei Mesinger for very useful comments, and is
grateful for support from the ICRR in Tokyo, Japan, the Moore
Distinguished Scholar program at Caltech, and the John Simon
Guggenheim Memorial Foundation, as well as Israel Science Foundation
grant 629/05.


\label{lastpage}

\end{document}